\documentclass[
  ,final            
  ]
  {aipproc}

\layoutstyle{8x11single}
\pdfoutput=1

\begin{document}

\title{LHCb detector status and commissioning}

\classification{29.40.Gx,29.40.Ka,29.40.Mc,29.40.Vj,13.20.He,14.65.Fy}
\keywords {LHCb detector status, LHCb commissioning, LHCb Higgs}

\author{B. Pietrzyk \\ (for the LHCb Collaboration)}{
  address={Laboratoire d'Annecy-le-Vieux de Physique des Particules LAPP,
IN2P3/CNRS, Universit\'e de Savoie, \\ F-74019 Annecy-le-Vieux
cedex, France }}

\begin{abstract}
 The LHCb detector status and commissioning is presented.
\end{abstract}

\maketitle


\section{Introduction}
LHCb is the Large Hadron Collider experiment for precise
measurement of CP violation and rare decays of beauty particles.
The two other LHCb talks will be presented later this week
\cite{cite:LHCbPresentations}. Therefore I will present only a
short introduction to the physics of LHCb before describing its
detector and commissioning.

\begin{figure}[h]
  \includegraphics[height=.3\textheight]{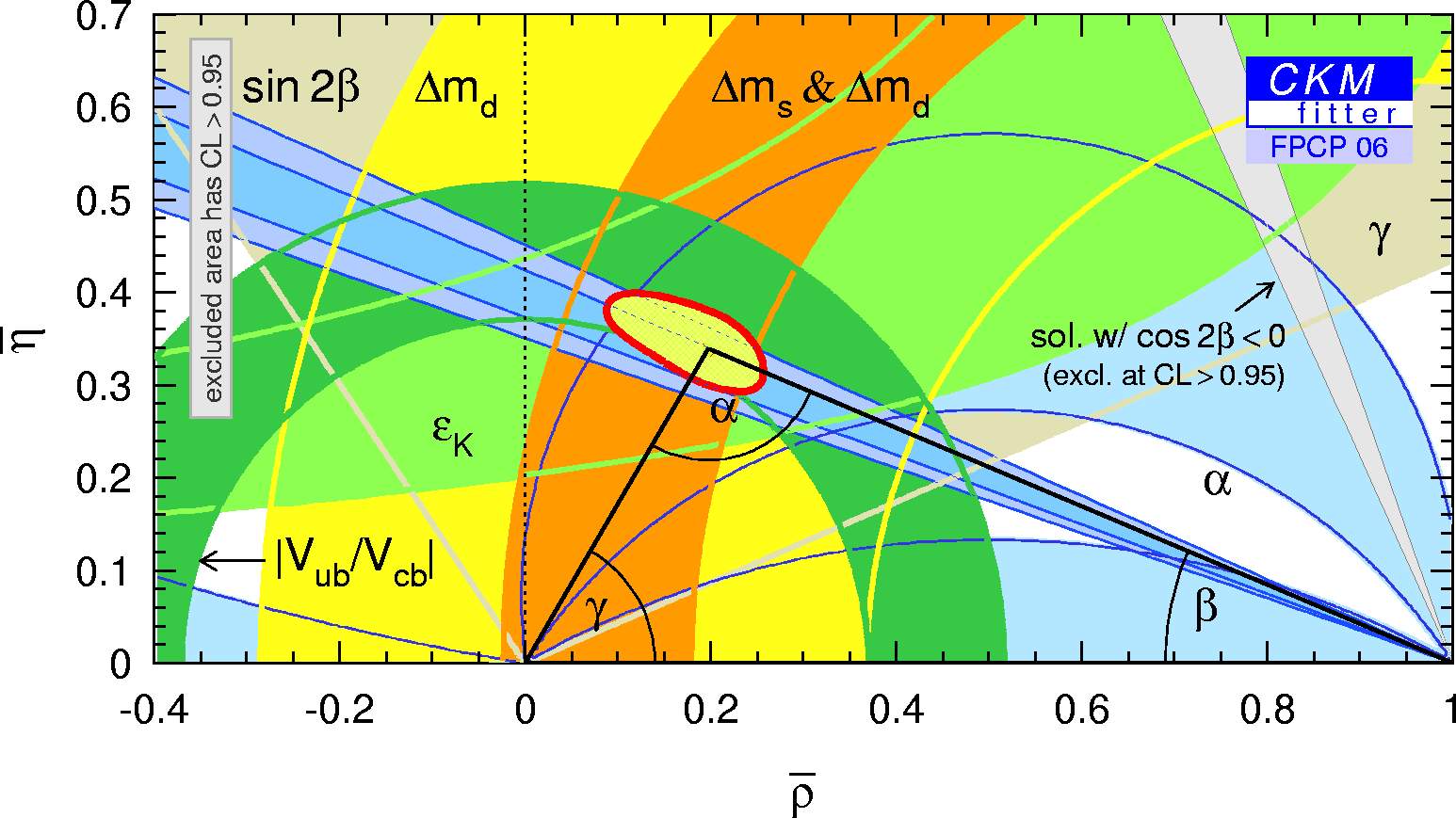}
  \caption{Combination of measurements obtained by the CKMfitter group.}
  \label{fig:CKMfitter}
\end{figure}

Excellent results have been obtained by the BELLE, BABAR, CDF and
D0 Collaborations on beauty physics. These measurements gave
coherent results as it can be seen in the combination obtained by
the CKMfitter group \cite{cite:CKMfitter} and presented on Fig.
\ref{fig:CKMfitter}. So far, there are no indications for New
Physics (NP).  However the effects of different NP models have
been predicted by many theorists and their non-observation results
in strong constraints on the parameters of these models.

\begin{figure}[b]
  \includegraphics[height=.3\textheight]{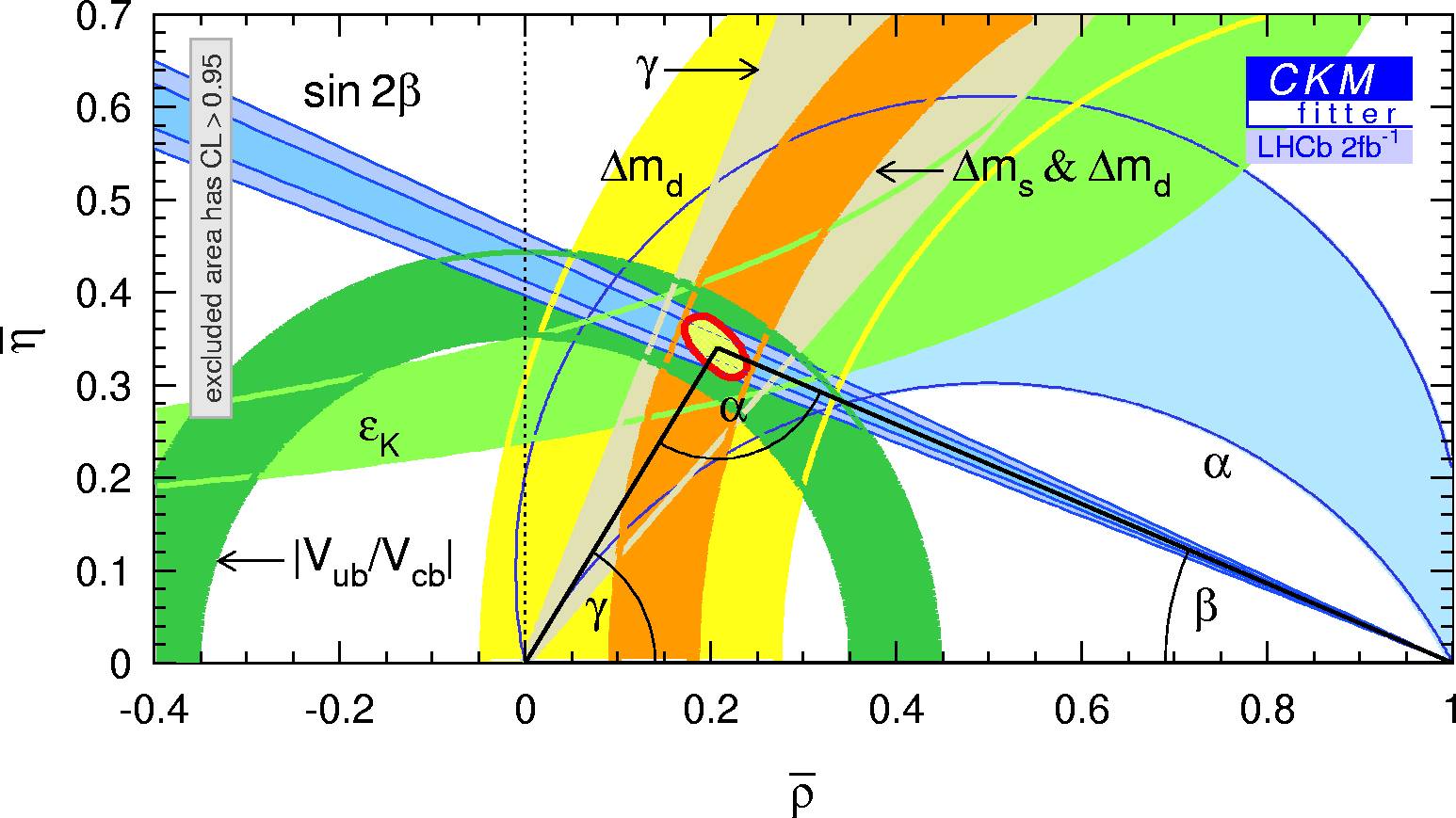}
  \caption{A possible CKMfitter combination using LHCb measurements
  with 2 fb$^{-1}$.}\label{fig:CKMfitter2fb}
\end{figure}

The three angles of CKM triangle are measured with the precision
presented in the Table. The $\gamma$ angle is still not measured
with significant precision. LHCb with 2 fb$^{-1}$, which is
nominal one year luminosity of LHCb, will measure many parameters
used in the CKMfitter compilation with higher precision and,
therefore, will give more constraints on NP. Particularly, the
three angles of CKM triangle will be measured with precision given
in the Table. The measurement of $\gamma$ will reach a precision
of 5$^\circ$. This angle will be measured in channels like
(B$_s\rightarrow$D$_s$K ) where the measurements are not effected
by the NP contribution and in channels where they are
(B$\rightarrow \pi \pi$, B$_s\rightarrow$KK,
B$\rightarrow$DK$^*$). Therefore LHCb will be able to determine
contribution of NP to these measurements.
Fig.~\ref{fig:CKMfitter2fb} shows how CKMfitter group compilation
could look like using LHCb measurements with 2 fb$^{-1}$.

\begin{figure}[t]
\begin{minipage}[t]{8cm}
 \includegraphics[height=.3\textheight]{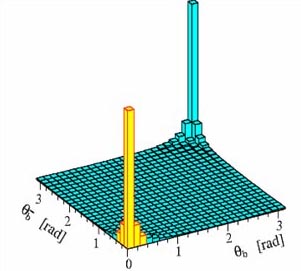}
\end{minipage}
\hfill
\begin{minipage}[t]{8.7cm}
\vspace*{-5cm}. \hspace*{2.5cm}TABLE \vspace*{0.5cm}
\\ \hspace*{1cm}
\begin{tabular}{lll} \hline
CKM & Current value& LHCb \\
angle & and precision & 2fb$^{-1}$ \\ \hline
$\beta$ & 21.7$^\circ$$^{+1.3}_{-1.2}$& $\pm0.5^\circ$ \\
$\alpha$ & 100$^\circ$$^{+15}_{-9}$& $\pm10^\circ$ \\
$\gamma$ & 63$^\circ$$^{+35}_{-25}$& $\pm5^\circ$\\\hline
\end{tabular}
 \end{minipage}
\caption{ (left) Angular distribution of b$\bar{\rm b}$ pair
production at the LHC; \hspace*{0.5cm} {\bf TABLE}\hspace*{0.3cm}
(right) Current precision of measurements of the three CKM angles
and the estimated precision using LHCb measurements with 2
fb$^{-1}$.} \label{fig:TheThe}
\end{figure}

The LHCb spectrometer will measure forward hadron production at
the pp collider. At the LHC the b$\bar{\rm b}$ pairs are produced
mostly in the forward direction as shown in Fig.~\ref{fig:TheThe}.
As at the Tevatron, different b hadrons are produced: B$_d$,
B$_u$, B$_s$, B$_c$, $\Lambda_b$, ... The b$\bar{\rm b}$
cross-section is $\sim$500$\mu$b and 10$^{12}$ b$\bar{\rm b}$
pairs/year (10$^7$s) reach the LHCb spectrometer. The LHCb
acceptance is more forward than the one of ATLAS/CMS and the
observed b$\bar{\rm b}$ cross-section is higher. The luminosity at
the LHCb interaction point is intentionally limited to
2$\times$10$^{32}$ cm$^{-2}$s$^{-1}$ in order to observe one
interaction per bunch crossing on average.

\section{Detector status}

For successful measurements of beauty physics the following
detector requirements are needed:
\begin{itemize}
\item Good triggering, to select interesting events from huge
background \item Good vertexing, to measure decay points and
reduce backgrounds \item Good tracking, to reconstruct tracks and
measure well their momenta \item Good particle identification, to
prevent decay products of one decay mode becoming the background
to another mode where kinematical separation is not sufficient
\item High speed DAQ coupled to large computing resources for data
processing
\end{itemize}

\begin{figure}[t]
  \includegraphics[height=.35\textheight]{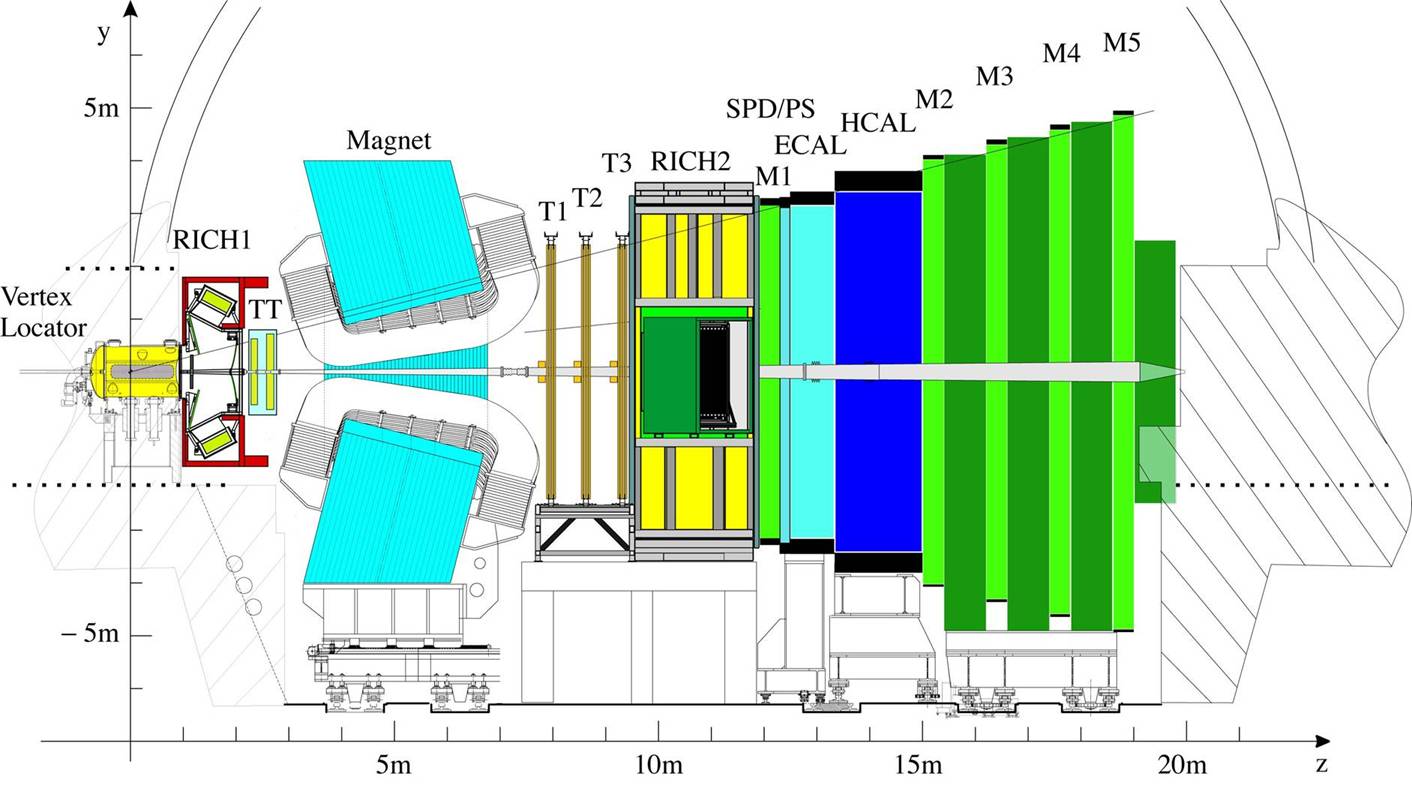}
  \caption{The LHCb detector.}\label{fig:LHCbDetector}
\end{figure}

The LHCb detector is presented in Fig.~\ref{fig:LHCbDetector}. The
protons collide inside the Vertex Locator. Products of interaction
are identified by two RICH detectors located on two sides of the
Magnet. The second RICH detector is followed by the calorimeter
detectors and moun spectrometer. The tracks are measured by
Trigger Tracker (TT) chambers and Tracking (T) stations located
before and after the Magnet. There is no material inside the
Magnet.

\begin{figure}[b]
\begin{minipage}[b]{8cm}
 \includegraphics[height=.19\textheight]{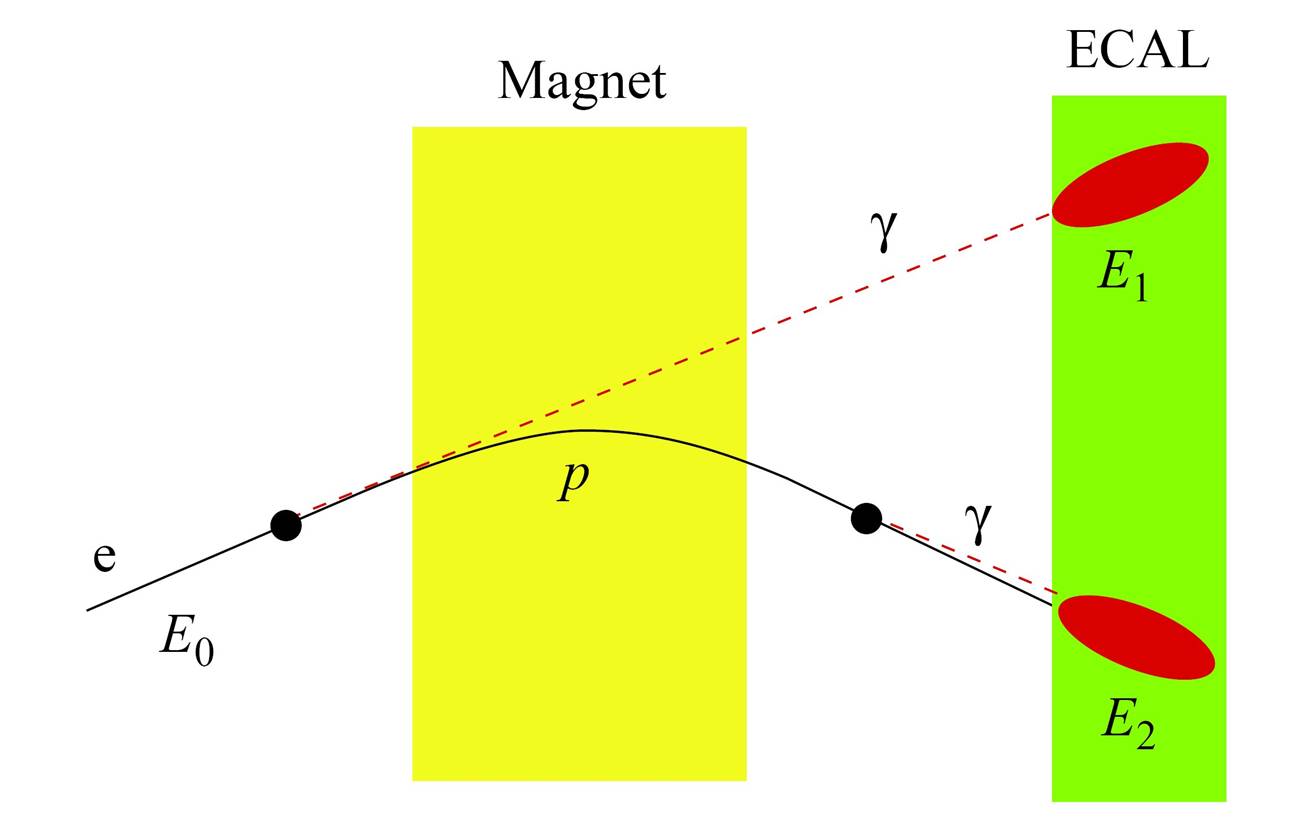}
\end{minipage}
\hfill
\begin{minipage}[b]{8.7cm}
\vspace*{-5cm} \hspace*{2.5cm}
 \includegraphics[height=.19\textheight]{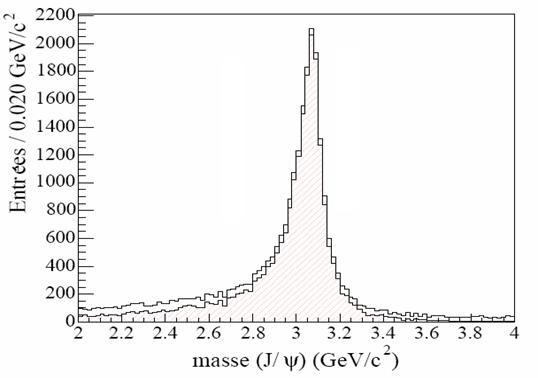}
\end{minipage}
  \caption{Photon radiation by electrons in LHCb (left) and the J/$\psi$
  reconstruction after adding radiated photons to electrons (right).}
  \label{fig:ElectronsJpsi}
\end{figure}

Importance of material absence inside the Magnet is best seen on
the example of electrons, particles which would be mostly affected
by its presence. In LHCb electrons radiate photons either before
or after the Magnet as shown in Fig.~\ref{fig:ElectronsJpsi}
(left). The clusters from the photons emitted by the electrons
after the Magnet are merged with those from electrons. Electrons
are identified by comparing energy measured in the electromagnetic
calorimeter ECAL with track momentum measured in the magnetic
field. The position of clusters from photons radiated by electrons
before the Magnet is precisely known by extrapolating the
direction of the corresponding electron tracks. J/$\psi
\rightarrow e^+e^-$ reconstruction is obtained by adding radiated
photons to electrons and is shown in Fig. \ref{fig:ElectronsJpsi}
(right).

Vertexing and tracking is assured by 21 Vertex Locator silicon
stations and by TT and T chamber stations. The track density is
high, therefore TT station and inner part (IT) of T stations is
made in silicon technology, the outer part (OT) is made in straw
tube technology. The excellent track reconstruction efficiency
>95\%, momentum resolution $\Delta$~p/p~$\sim 0.4$\%, impact
parameter resolution $\sim 20 \mu$m and proper time resolution
$\sim$ 40 fs are obtained. The fraction of false tracks "invented"
by the reconstruction program is low $\sim$ 4\%.

The particles are identified by two RICH detectors. RICH1
identifies low momentum tracks. using two radiators, Aerogel and
C$_4$F$_{10}$. CF$_4$ allows to identify high momentum tracks in
RICH2. The kaons are identified with 88\% efficiency and a
corresponding misidentification rate of pions as kaons is 3\%.
There is better than 3$\sigma$ separation between pions and kaons
with momenta between 3 and 80 GeV.

\begin{figure}[t]
  \includegraphics[height=.5\textheight]{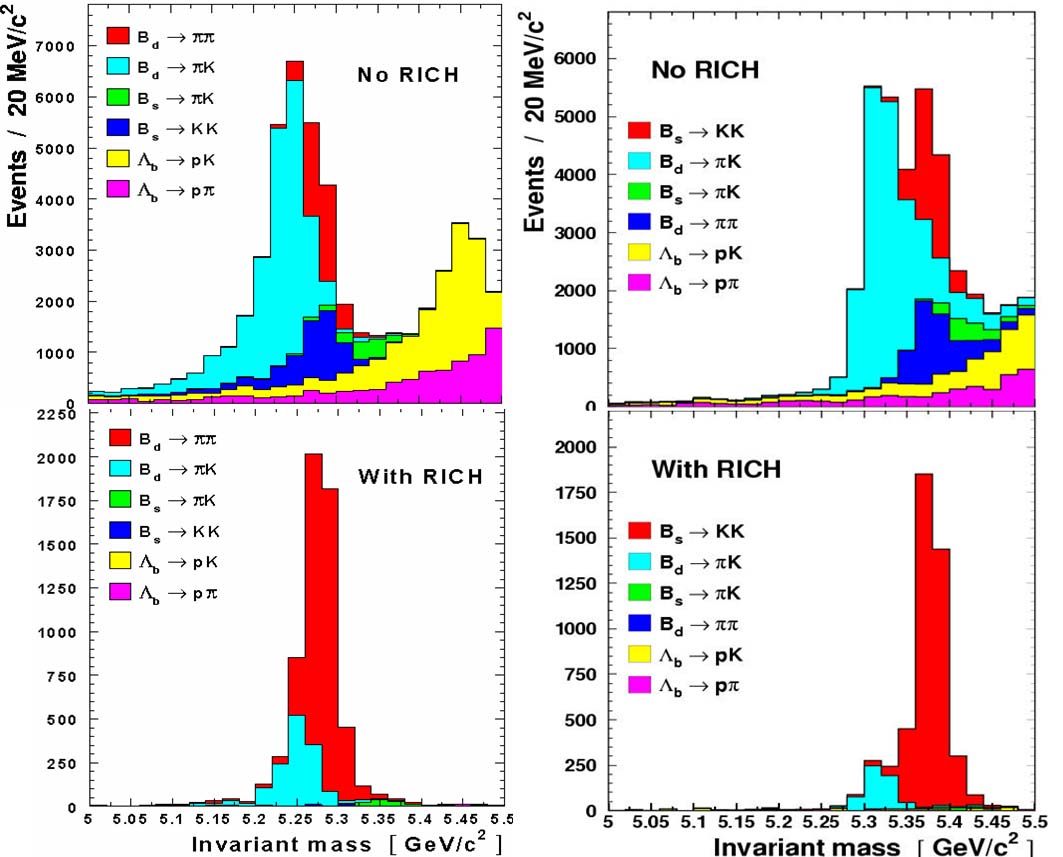}
  \caption{B$_d\rightarrow \pi \pi$ and B$_s\rightarrow$KK
  selection with and without RICH identification.}\label{fig:BsBd}
\end{figure}

The importance of good particle identification in b physics is
shown in an example of B$_s \rightarrow$ KK and B$_d \rightarrow
\pi \pi$ selection presented in Fig.~\ref{fig:BsBd}. Without RICH
identification the two selections are strongly contaminated by
background. The RICH detectors allow a very clean selection of
these two decays. This is a unique feature at hadron colliders.

The Hardware Trigger is reducing the 40 MHz bunch crossing rate
into 1 MHz rate transmitted to the software trigger. High p${\rm
_T}$ electrons, photons and hadrons measured by the Calorimetry
and high p${\rm _T}$ muons and muon pairs measured by the Muon
Spectrometer are selected. Multiple interactions are rejected by
the Pile-up system located near the Vertex Locator.

The Software Trigger is using the data sent from different
subdetectors through the Readout Network to CPU processors. Full
detector information is therefore available at 1 MHz. The hardware
trigger is confirmed or not and then more information is used
step-by-step and uninteresting events are rejected. Ultimately,
data are stored at a rate of 2 kHz.

\begin{figure}[t]
  \includegraphics[height=.35\textheight]{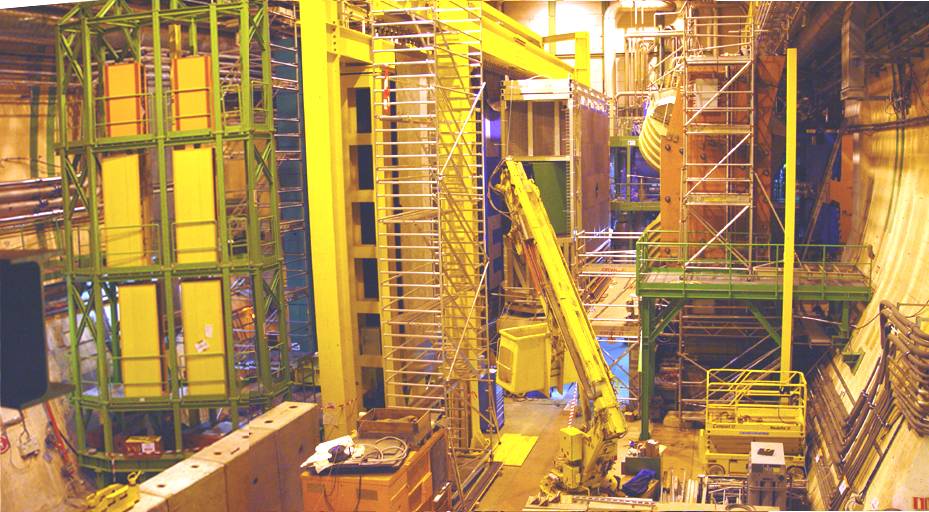}
  \caption{LHCb detector in spring 2006.}\label{fig:PhotoPit}
\end{figure}

Fig.~\ref{fig:PhotoPit} shows the LHCb detector construction
status in spring 2006. The interaction point is located inside the
LHC tunnel on the right hand side of the Figure. The particles
produced at the interaction traverse then, from the right to the
left hand side, the RICH1 and the Magnet. The three tracking
chamber stations will be installed in the empty space after the
Magnet. RICH2 detector is already installed followed by the
Calorimeter detectors. ECAL and HCAL detectors are already in
place as well as the thin wall of lead converter. On both sides of
the lead the thin scintillator pad SPD and PRS detectors will be
installed in order to measure the beginning of shower development
and distinguish between e's and $\gamma$'s. The ECAL and the HCAL
detectors are placed on the chariots and can be open and closed in
any possible configuration. The SPD, PRS and lead wall are fixed
from above and similarly can be opened and closed in any possible
configuration. The muon detector wall is seen on the left hand
side of Fig.~\ref{fig:PhotoPit}.

The current status and construction planning of the LHCb detector
is given below:
\begin{itemize}
\item{Magnetic field has been already successfully mapped.}
\item{Tracking chambers: OT production is finished, IT and TT
chambers are being produced, installation in the pit will end in
Autumn 2006.} \item{Calorimetry: SPD and PRS detectors will be
installed in Summer 2006.} \item{Vertex detector: vacuum tank is
already installed, silicon modules are produced and tested now and
will be installed in the pit on the beginning of 2007.} \item{Beam
pipe is composed of three sections in beryllium and the last one
in stainless steel, installation starts soon and will be completed
at the end of 2006.} \item{RICH1: shielding is already in place,
end of installation is planned for the beginning 2007.} \item{Muon
system: muon filter is in place, muon chambers are being produced,
installation has started already and will end at the beginning of
2007.} \item{Electronics: most of electronics will be installed at
the pit in the Summer and Autumn 2006, the tests within
subdetectors will continue till the end of 2006, when the global
commissioning will start.}
\end{itemize}

\section{Global commissioning}
The global commissioning without beam will be made in the first
half of 2007. The control and safety will be commissioned, the DAQ
and the electronic calibration procedures will be tested. The
scalability of the system will be checked and improved when
needed.

Circulated beam will be available in the Summer of 2007. This is
important since LHCb is a forward detector and therefore cosmic
rays (mostly vertical) are not very useful. On the other hand
interactions of LHC beam with beam gas give useful tracks for time
and position alignment.

During the pilot run the alignment will be measured without
magnetic field and finally trigger setup and data taking will be
made with magnetic field.

\section{Bonus: light Higgs search}

\begin{figure}[t]
  \includegraphics[height=.35\textheight]{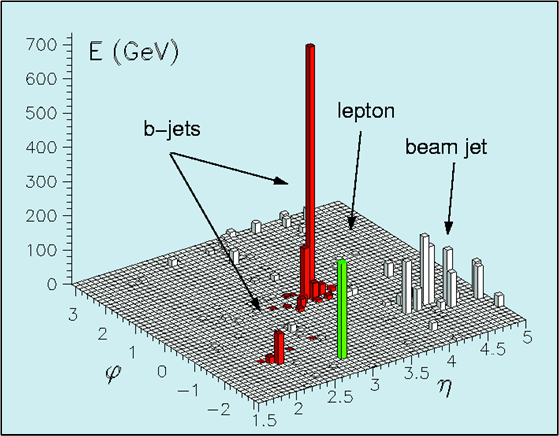}
  \caption{A possible signature of light Higgs event in LHCb.}\label{fig:Higgs}
\end{figure}

At LHC a significant fraction ($\sim$ 30\%) of the light Higgs
bosons, currently being search for at the Tevatron, are emitted
forward within the acceptance of the excellent LHCb spectrometer.
LHCb has very good b-quark identification and its spectrometer
will be very well calibrated with the large number of B meson
peaks. The LHCb Collaboration is investigating a possibility of
Higgs discovery by measuring Higgs decay into b$\bar{\rm b}$ jets
associated with high p${\rm _T}$ lepton in order to reduce high
t$\bar{\rm t}$ production background. A typical light Higgs
production event is shown in Fig.~\ref{fig:Higgs}. The sensitivity
to such light Higgs events is currently under study.

\section{Final remarks}

The LHCb detector construction and commissioning is progressing
efficiently. It will be ready in 2007 to observe first collisions
at LHC and soon after to get first physics results. More details
can be found in \cite{cite:LHCb} and
\cite{cite:LHCbPresentations}.

I would like to thank the organizers of this Symposium for the
excellent organization, many members of the LHCb collaboration for
their help in preparation of this presentation and Stephane
T'Jampens for CKMfitter plots.

\bibliographystyle{aipproc}   

\bibliography{sample}

\IfFileExists{\jobname.bbl}{}
 {\typeout{}
  \typeout{******************************************}
  \typeout{** Please run "bibtex \jobname" to optain}
  \typeout{** the bibliography and then re-run LaTeX}
  \typeout{** twice to fix the references!}
  \typeout{******************************************}
  \typeout{}
 }


\end{document}